\documentclass{article}
\usepackage{dn_report,graphicx,rotating}
\newcommand{\Pt}{P_{\tau}}

\begin{document}

\make_cover_page{TRI-DN-05-7}{R. Baartman} {Electrostatic Bender Optics}{The
relativistically-correct Hamiltonian and transfer matrix of electrostatic
benders is derived. This is the general case where the bender electrodes
have curvature in the non-bend direction.}

\section{Introduction}
A common approach, used especially by those accustomed to only magnetic
elements, is to let the third momentum be $\Delta p/p$. This is not the
simplest approach, since when electric fields are included, it is not
conserved. This means that when a particle enters the electrostatic element
off-axis, it must receive a ``kick'' to get into the potential
field. This kicks $p$, but to first order leaves $E$ unchanged. For an
electrostatic bend of radius $A$, this kick is $\Delta p/p=\pm x/A$; the
upper sign is for entry, the lower for exit.

A better approach is therefore to let the third momentum be $E$. 
Since the fields are static, $E$ is conserved; no
kicks are required.

For independent variable $s$, on a reference trajectory curving in the
$xs$-plane with curvature $h=1/A$, the Hamiltonian $H=-p_s$ is well-known, and I
will not derive it here:
\begin{equation}H=-(1+hx)\sqrt{\left({E-q\Phi\over c}\right)^2-m^2c^2-p_x^2-p_y^2}\end{equation}
We write $E=E_0+\Delta E$, and note that the large quantity under the
square root sign is 
\begin{equation}p_0^2=E_0^2/c^2-m^2c^2=(\gamma^2-1)m^2c^2=(\beta\gamma mc)^2,\end{equation} the
square of the reference momentum.

\begin{equation}H=-(1+hx)p_0\sqrt{1+{2E_0(\Delta E-q\Phi)\over p_0^2c^2}+\left({\Delta
      E-q\Phi\over p_0c}\right)^2-{p_x^2\over p_0^2}-{p_y^2\over p_0^2}}\end{equation}

Let us transform so that the third coordinate is not time, but a relative
distance deviation w.r.t.\ the reference particle: i.e.\ from $(t,-\Delta
E)$ to $(\tau,p_\tau)$ where $\tau\equiv s-\beta ct$, $p_\tau=\Delta
E/(\beta c)$. The generating function is 
\begin{equation}F(t,p_\tau)=(s-\beta ct)p_\tau\end{equation} then the new Hamiltonian is
$K=H+\partial F/\partial s=H+p_\tau$
\begin{equation}K=p_\tau-(1+hx)p_0\sqrt{1+2\left({p_\tau\over
      p_0}-{q\Phi\over\beta cp_0}\right)+\beta^2\left({p_\tau\over p_0}-{q\Phi\over\beta cp_0}\right)^2-{p_x^2\over
    p_0^2}-{p_y^2\over p_0^2}}\end{equation}
This cleans up considerably if we scale all momenta and the Hamiltonian by
$p_0$: $P_x=p_x/p_0$, $P_y=p_y/p_0$, $P_\tau=p_\tau/p_0$,
$\tilde{K}=K/p_0$, and introduce the scaled potential $V={q\Phi\over\beta cp_0}$:
\begin{equation}\tilde{K}=
P_\tau-(1+hx)\sqrt{1+2\left(P_\tau-V\right)+\beta^2\left(P_\tau-V\right)^2-P_x^2-P_y^2}\end{equation}
This also makes the momenta accord with the more usual definitions, since,
as we will see, $P_x=x'$, $P_y=y'$, and outside the electric field,
$P_\tau=\Delta p/p$.

\section{Potential}
To second order, the potential $V$ is given by 
\begin{equation}\label{pot}V=hx-h(h+k){x^2\over 2}+hk{y^2\over 2},\end{equation} where $h=1/A$ and $k=1/A_y$ is the curvature in
the non-bend direction. This can be derived by solving Laplace's equation
in a curvilinear coordinate system that has different curvatures in the
$xs$- and $xy$-planes, then transforming back to the chosen dynamical
system, which is curved only in the $xs$-plane \cite{snowmass}. We can,
however, show the above potential to be correct for the two simplest cases
-- cylindrical and spherical.

\begin{quote}{\bf Example: Cylindrical bend}

  Here $k=0$, the potential is $V=-\log (A/r)=\log(1+hx)$, since the
  distance $r$ to the bend centre is $x+A$. Expanding, we find
\begin{equation}V=hx-{h^2x^2\over 2}+{h^3x^3\over 3}-\cdots\end{equation}
This agrees with eqn.\,\ref{pot} for $k=0$.
\end{quote}

\begin{quote}{\bf Example: Spherical bend}

Here $k=h$, the potential is
$V=1-A/r=1-1/\sqrt{1+2hx+h^2x^2+h^2y^2}$, since the distance $r$ to the
bend centre is $\sqrt{(x+A)^2+y^2}$. Expanding, we find
\begin{equation}V=hx-h^2x^2+{h^2y^2\over 2}+h^3x^3-{3h^3xy^2\over 2}+\cdots\end{equation}This agrees with eqn.\,\ref{pot} for $k=h$.
\end{quote}

In any case, the first term, needed to ensure that the reference
trajectory is $x=0$, yields the required electric field on the reference
orbit:
\begin{equation}{\cal E}= -{\partial\Phi\over\partial x}=-{\beta cp_0\over
  q}\,\left.{\partial V\over\partial x}\right|_{x=0}={\beta^2\over A}{E_0\over q}.\end{equation}
In the non-relativistic limit, the electric field is twice the beam kinetic
energy divided by charge and bend radius: $q{\cal E}=mv^2/A$.

\section{Hamiltonian and Transfer Matrix}
The first order terms in the resulting Hamiltonian all cancel, so when
expanded to second order it is
\begin{equation}\tilde{K}={P_x^2\over 2}+{P_y^2\over 2}+{\Pt^2\over 2\gamma^2}-{2-\beta^2\over A}\,x\Pt+
{\xi^2\over 2A^2}\,x^2+{\eta^2\over 2A^2}\,y^2\end{equation}
The parameters $\xi$ and $\eta$ are introduced as they parameterize the $x$
and $y$ focusing strengths: $\xi^2+\eta^2=2-\beta^2, \eta^2=k/h=A/A_y$, $A_y$ being the
curvature radius in the non-bend direction. In the non-relativistic limit, 
for a cylindrical bend, $\xi=\sqrt{2},\eta=0$; for a spherical bend, $\xi=\eta=1$.

The transfer matrix is easily derived from this Hamiltonian $\tilde{K}$:
\renewcommand{\arraystretch}{1.5}
\begin{equation}
\left( \begin{array}{cccccc}   
\cos \xi\theta & {A\over \xi}\sin \xi\theta & 0 & 0 & 0 & {2-\beta^2\over \xi^2}A(1-\cos \xi\theta) \\
-{\xi\over A}\sin \xi\theta & \cos \xi\theta & 0 & 0 & 0 & {2-\beta^2\over \xi}\sin \xi\theta \\
          0 & 0 & \cos \eta\theta & {A\over \eta}\sin \eta\theta & 0 & 0 \\
0 & 0 & -{\eta\over A}\sin \eta\theta & \cos \eta\theta & 0 & 0 \\
-{2-\beta^2\over \xi}\sin \xi\theta & -{2-\beta^2\over \xi^2}A(1-\cos \xi\theta) & 0 & 0 & 1 &
A\theta\left[{1\over\gamma^2}-{\left(2-\beta^2\over \xi\right)^2}\left(1-{\sin \xi\theta\over \xi\theta}\right)\right] \\
          0 & 0 & 0 & 0 & 0 & 1 
\end{array}\right)\nonumber
\end{equation}
\renewcommand{\arraystretch}{1} Interestingly, in the extreme relativistic
limit ($\beta=1$), this is identical to the matrix for a magnetic bend with
field index.

\end{document}